\documentclass[aps,%
pre,%
 twocolumn,%
floatfix]{revtex4}
 \usepackage{epsfig} 
\usepackage{here}
\newcommand{\IGN}[1]{}
\begin{document}
%\twocolumn

\IGN{ % e-print info header.
\\
Title: Stability borders of feedback control of delayed measured systems
Authors: Jens Christian Claussen
 (Theoretical Physics Kiel, Germany)
Comments: 5 pages LaTeX using revtex and epsfig (4 figs)
Report-no: ITAP-Claussen-98b Sep7(1998) with minor changes
Subj-class: Chaotic Dynamics
\\
When stabilization of unstable periodic orbits or fixed points by the 
method given by Ott, Grebogi and Yorke (OGY) has to be based on a 
measurement  delayed by $\tau$ orbit lengths, the performance 
of unmodified OGY method is expected to decline. For experimental 
considerations, it is desired to know the range of stability with 
minimal knowledge of the system. We find that unmodified OGY control fails 
beyond a maximal Ljapunov number of $\lambda_{max}=1+\frac{1}{\tau}$.
In this paper the area of stability is investigated both for OGY control
of known fixed points and for difference control of unknown or inaccurately 
known fixed points. An estimated value of the control gain is given.
Finally we outline what extensions have to be considered if one wants to 
stabilize fixed points with Ljapunov numbers above $\lambda_{max}$.
\\
}

\title{Stability borders of feedback control of delayed measured
  systems}
\author{Jens Christian Claussen}
%\address{
\affiliation{Institut f\"ur Theoretische Physik und Astrophysik der Universit\"at Kiel,
24098 Kiel, Germany
%\date{September 7, 1998, minor changes \today}
}
\date{April 15, 2002, revised May 21,2004}

%\twocolumn

\begin{abstract}                % DON'T CHANGE THIS LINE
When stabilization of unstable periodic orbits or fixed points by the method
given by Ott, Grebogi and Yorke (OGY) has to be based on a 
measurement delayed by $\tau$ orbit lengths, the performance 
of unmodified OGY method is expected to decline.
For experimental considerations, it is desired to know the range
of stability with minimal knowledge of the system.
We find that unmodified OGY control fails beyond a maximal Ljapunov number of
 $\lambda_{\rm{}max}=1+\frac{1}{\tau}$.
In this paper the area of stability is investigated both for OGY control
of known fixed points and for difference control of unknown 
or inaccurately known fixed points.
An estimated value of the control gain is given.
Finally we outline what extensions have to be considered
if one wants to stabilize fixed points with Ljapunov numbers above
$\lambda_{\rm{}max}$.
\end{abstract}
\maketitle

\section{Introduction}
The appearence of
delay is a common problem in the control of chaotic
systems.
The effective delay time in any feedback loop is the sum of at least 
three delay times, the time of measurement, the time to compute 
the appropriate control amplitude, and the response time of the
system to the applied control.
If the applied control additionally has to propagate through the system
\cite{mausbach95},
these response time may extend to one or more cycle lengths.

In this paper we investigate time-discrete systems 
and focus on the question what limitations occur if one
applies the control method given by Ott, Grebogi and Yorke (OGY) \cite{ogy90}
or difference feedback \cite{bielawski93a}
in the presence of time delay.

Stabilization of chaotic systems by small perturbations 
in system variables \cite{hubler} 
or control parameters  \cite{ogy90}
has become a widely discussed topic with applications
in a broad area from technical to biological systems.
The OGY method given by Ott, Grebogi and Yorke
\cite{ogy90}
stabilizes unstable fixed points (or unstable periodic orbits
utilizing a Poincar\'e surface of section \cite{schusterbuch}) 
by feedback that is applied in vicinity of the fixed point $x^*$ 
of a discrete dynamics $x_{t+1}=f(x_t,r)$.
The amplitude of the feedback $r_t=r-r_0$ 
added to the control parameter $r_0$ 
is proportional 
(with some user-adjustable parameter $\varepsilon$ determining the strength 
of control)
to the distance
$x-x^*$ from the fixed point, 
\begin{eqnarray}
r_t = \varepsilon (x_t-x^{*}),
\end{eqnarray}  
and the feedback gain
can be determined from a linearization around the fixed point:
Neglecting higher order terms, we have
\begin{eqnarray}
\!f(x_t,r_0+r_t)\!&=&\! f(x^{*},r_0)
+ (x_t-x^{*}) \cdot
\left(\frac{\partial{} f}{\partial{} x}\right)_{x^{*},r_0}
   \nonumber\\&&
+r_t \cdot 
 \left(\frac{\partial{} f}{\partial{} r}\right)_{x^{*},r_0}
\nonumber
\\
&=& f(x^{*},r_0)+ \lambda (x_t-x^{*})+\mu r_t
\nonumber  
\\
&=& f(x^{*},r_0)+ (\lambda + \mu \varepsilon)\cdot (x_t-x^{*})
\end{eqnarray}  
where the Taylor coefficients $\lambda$ and $\mu$ are
fixed values for each orbit.
The second expression vanishes for $\varepsilon=-\lambda/\mu$,
that is, in linear approximation the system arrives exactly
at the fixed point in the next time step, $x_{t+1}=x^{*}$.

As the uncontrolled system at hand is assumed to be unstable in the fixed point,
we generally have the situation $|\lambda|>1$.
The system with applied control is stable, in linear approximation, if the 
absolute value of the eigenvalues of the iterated map
is smaller than one,
\begin{eqnarray}
|x_{t+1}-x^{*}|=
|(\lambda + \mu \varepsilon)\cdot (x_{t}-x^{*})|
<|x_{t}-x^{*}|,
\end{eqnarray} 
i.~e. $|(\lambda + \mu \varepsilon)|<1$. 
Therefore $\varepsilon$ has to be chosen 
between $(-1-\lambda)/\mu$ and $(+1-\lambda)/\mu$,
and this interval is of width $2/\mu$ and independent of $\lambda$.
For OGY control the range in $\lambda$ that can be controlled 
remarkably is not
bounded, which will appear to be different for delayed measurement
(see next section).

It should be mentioned that the stabiliy analysis of the one-dimensional case
holds also for higher-dimensional
systems provided there is only one unstable direction.
One can transform on the eigensystem of the Jacobi matrix 
$\frac{\partial{} f}{\partial{} r}$
and finds again the equations of the one-dimensional case,
reflecting that one only needs to apply control in the unstable direction
(see e.g. \cite{claussen98c,claussenthesis}).

\section{Delayed control of iterated maps}
We want to know what limitations occur if the OGY rule
is applied without modification.
In OGY control, the control parameter $r_t$ is time-dependent,
and without loss of generality we assume that $x^*=0$ and
that $r_t=0$ if no control is applied.
For simplicity, we discuss the case $\tau=1$ first.
For one time step delay, 
instead of $r_t = \varepsilon x_{t}$
we have the proportional feedback rule:
\begin{eqnarray}
r_t = \varepsilon x_{t-1}.
\end{eqnarray}  
Using the time-delayed coordinates $(x_t,x_{t-1})$,
the linearized dynamics of the system with
applied control
is given by
\begin{eqnarray}
\left(
\begin{array}{c}
x_{t+1}\\x_{t}
\end{array}
\right)
= 
\left(\begin{array}{cc}\lambda & \mu\varepsilon\\1 & 0\end{array}\right)
\left(
\begin{array}{c} 
x_{t}\\x_{t-1}
\end{array}
\right).
\end{eqnarray}

\noindent
The eigenvalues of 
\footnotesize
$\left(\begin{array}{cc}\lambda & \mu\varepsilon\\1 & 0\end{array}\right)$
\normalsize
%%% $\alpha^2-\lambda\alpha-\varepsilon\mu=0$
are given by
%\begin{eqnarray}
$
\alpha_{1,2}=\frac{\lambda}{2}
\pm \sqrt{\frac{\lambda^2}{4}+\varepsilon\mu}.
$
%\end{eqnarray}
Hence control can be achieved 
with $\varepsilon$ being \typeout{CHANGED}
in an interval 
$]-1/\mu,(1-\lambda)/\mu[$ with the 
width $(2-\lambda)/\mu$.

In contrast to the not-delayed case, we have a requirement
$\lambda<2$ for the Lyapunov number, i.~e. the direct application
of the OGY method fails for systems with a Lyapunov number 
of 2 and higher.
This limitation is caused by the additional degree of freedom
introduced in the system due to the time delay.

Now we consider the general case.
If the system is measured delayed by $\tau$ steps,
%\begin{eqnarray}
$
r_t = \varepsilon x_{t-\tau},
$
%\end{eqnarray}  
we write the dynamics in delayed coordinates
%%$\vec{y}(t) := $
$(x_t, x_{t-1}, x_{t-2},
\ldots x_{t-\tau})^{\mbox{\scriptsize\rm{}T}}$:
\begin{eqnarray}
\left(
\begin{array}{c}
x_{t+1}\\ \vdots\\\\\\\\\\\vdots\\x_{t-\tau+1}
\end{array}\right)
= 
\left(
\begin{array}{ccccccc}
\lambda&0&\cdots&&\cdots&0&\varepsilon\mu\\
1&0&&&&&0\\
0&1&\ddots&&&&\vdots\\
\vdots&&\ddots&&&&\\  
&&&&\ddots&&\\
\vdots&&&&\ddots&0&\vdots\\
0&\cdots&&\cdots&0&1&0
\end{array}
\right)
\left(
\begin{array}{c}
x_{t}\\ \vdots\\\\\\\\\\\vdots\\x_{t-\tau}
\end{array}\right)
\nonumber
\end{eqnarray}
The characteristic polynomial is given by 
(we define rescaled coordinates $\tilde{\alpha} := \alpha/\lambda$
and $\tilde{\varepsilon} = \varepsilon\mu/\lambda^{\tau+1}$)
\begin{eqnarray}
0&=&P(\alpha)=(\alpha-\lambda)\alpha^{\tau}-\varepsilon\mu
~~~~~~~~
\nonumber \\ 
\mbox{or}
~~~~~~~~
0&=&P(\tilde{\alpha})=(\tilde{\alpha}-1)\tilde{\alpha}^{\tau}
-\tilde{\varepsilon}.
\label{eq:charpoly}
\end{eqnarray}

\begin{figure}[htbp]
\begin{center}
\noindent
\epsfig{file=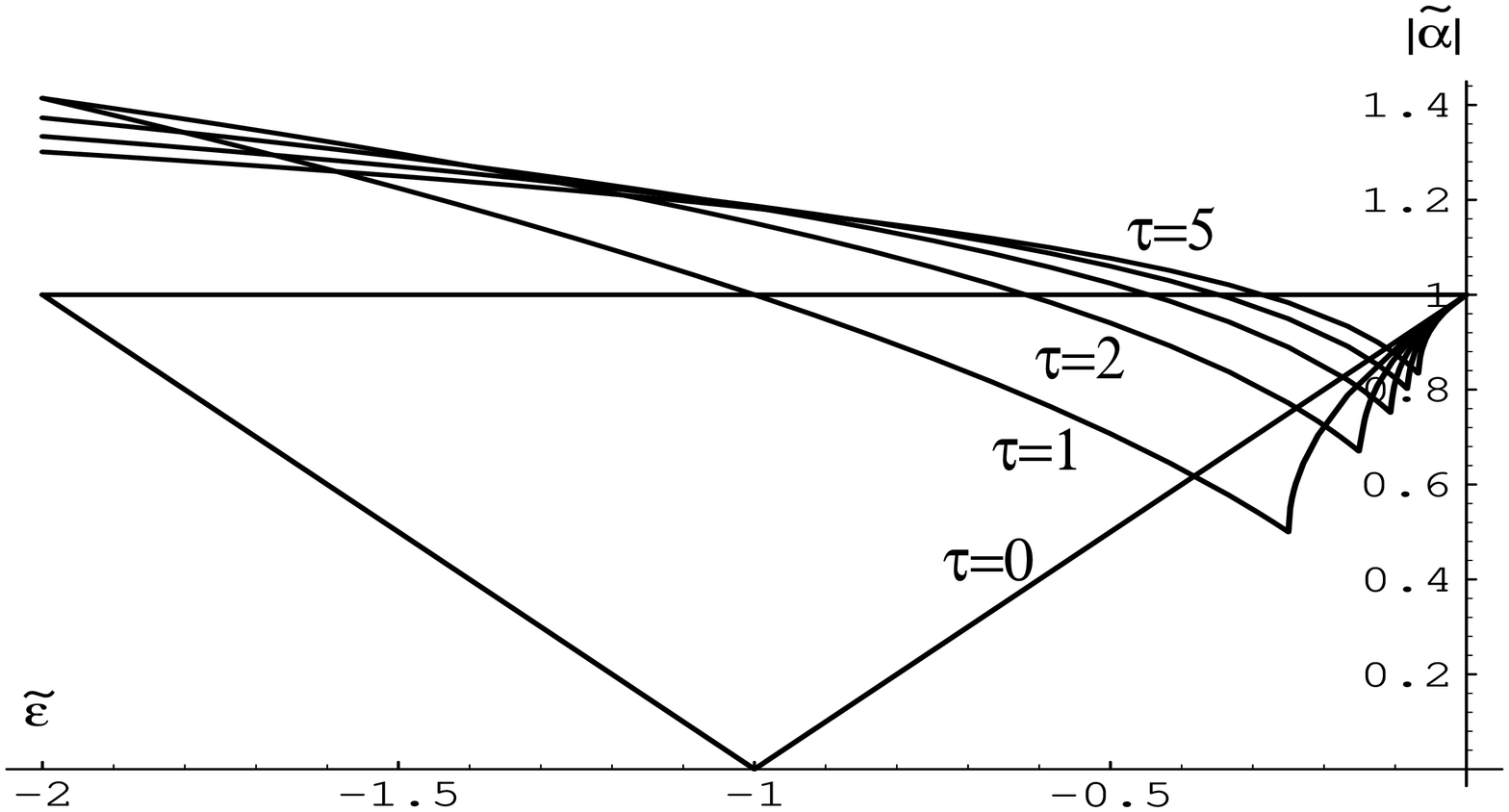, angle=0, width=0.8\columnwidth}
\end{center}
\caption[Verschiedene Verz\"ogerungen]{
Control intervals for several time delays $\tau=0\ldots{}5$:
The plots show the maximal absolute value of the eigenvalues
as a function of the rescaled control gain $\tilde{\varepsilon}$.
Values of $|\tilde{\alpha}|=1/\lambda$ correspond to 
$|\alpha|=1$ in (\protect\ref{eq:charpoly})
without rescaling, so one can obtain the range 
$]\varepsilon_-,\varepsilon_+[$ for which control is successfully acheived.
\label{fig:kurven4}}
\end{figure}

\noindent
Fig.~\ref{fig:kurven4} shows the maximum of the
absolute value of the eigenvalues, for $\tau=0,1,\ldots,5$.
In rescaled coordinates $\tilde{\alpha}=1/\lambda$ corresponds to
a control interval $\tilde{\varepsilon}_{\pm}(\tau,\lambda)$.
For
\begin{eqnarray}
\lambda_{\rm max} = 1 + \frac{1}{\tau}
\label{lmaxvontau}
\label{eq:lambdamaxogy}
\end{eqnarray}
the control interval vanishes, and for $\lambda\geq\lambda_{\rm max}(\tau)$
no control is possible.
Equation~(\ref{eq:lambdamaxogy}) and the subsequently derived stability
diagrams are the main result of this paper and are transferred to difference
control in section~\ref{sec:diffkontr}.

If we look at the Lyapunov exponent $\Lambda:=\ln \lambda$
instead of the Lyapunov number,
we find with $\ln x < (x-1)$ the inequality
\begin{eqnarray}
\Lambda_{\rm max} \cdot \tau < 1.
\label{eq:controllability}
\end{eqnarray}
Therefore, delay time and Lyapunov exponent limit each other 
if the system is to be controlled.
This is consistent with the loss of knowledge in the system
by exponential separation of trajectories.

\section{Stability diagrams by the Jury criterion}
For small $\tau$ one can derive easily the borders of the
stability area with help of the Jury criterion~\cite{jury}
(see Appendix~\ref{appj0}).
For $\tau=1$, the Jury coefficients are given by
$\alpha_1 = -{\lambda}/{(1+\mu\varepsilon)}$ 
and $\alpha_2 = -\mu\varepsilon$,
\IGN{
\begin{eqnarray}
\alpha_2 = -\mu\varepsilon 
~~~~~~~~
%\mbox{and}
~~~~~~~~
\alpha_1 = -\frac{\lambda}{1+\mu\varepsilon}
\end{eqnarray}
}%IGN
and for $\tau=2$ to $\tau=4$ the 
corresponding expressions are shown in
Appendix~\ref{appj1}.

\begin{figure}[htbp]
\begin{center}
\noindent
\epsfig{file=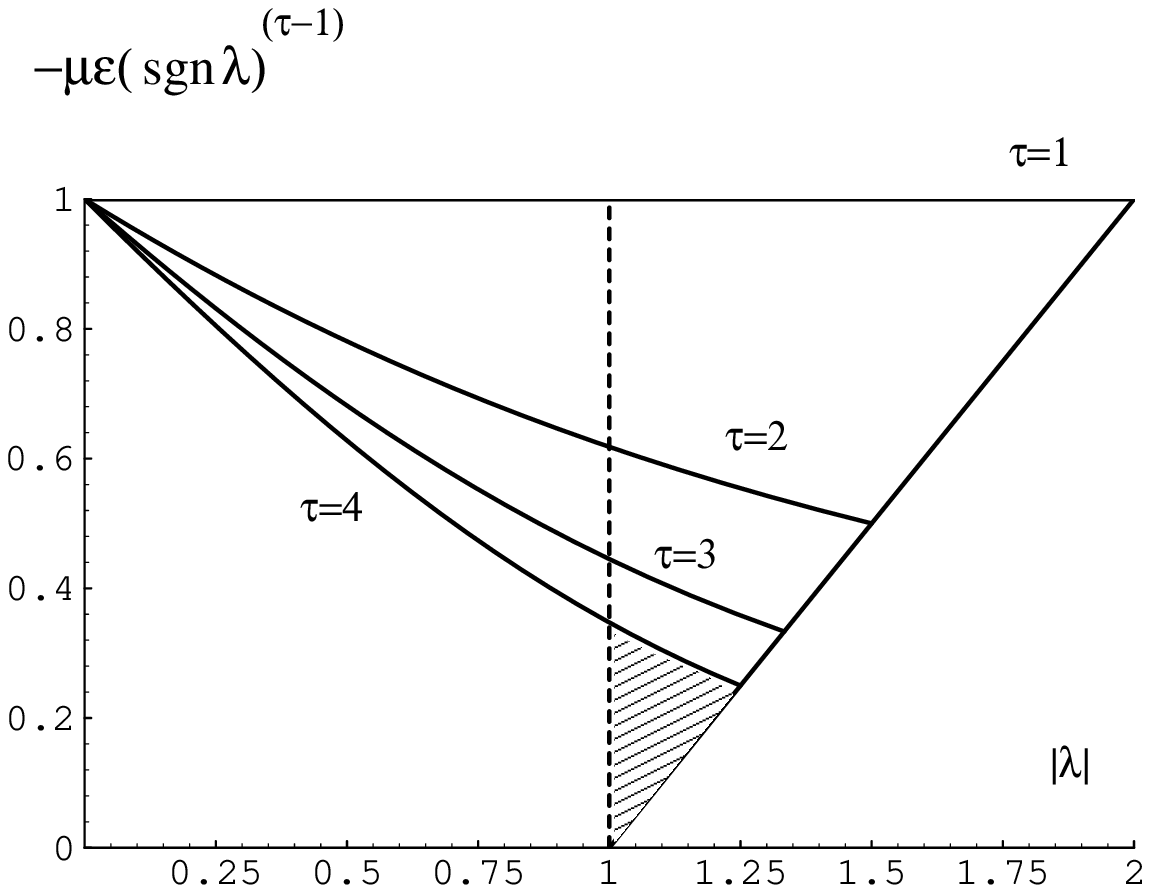, angle=0,width=0.9\columnwidth}
\end{center}
\caption[quadrant]{
Stability areas for $\tau=1, 2, 3, 4$, combined.
Only for $|\lambda|>1$ control is necessary (dashed line),
and the stability area 
(shaded for $\tau=4$)
extends to 
$|\lambda_{\rm{}max}|= 2, \, 3/2, \, 4/3, \, 5/4$.
%%%(ordinata: $0\leq|\lambda|\leq +2$ and abscissa: $0\leq|\mu\varepsilon|\leq +1$).
Note that still both positive and negative $\lambda$ can 
be controlled.
The abscissa $-\mu\varepsilon({\rm sgn}\lambda)^{(\tau-1)}$
takes into account that for odd $\tau$ a negative
$\mu\varepsilon$ is required, independent of the sign of $\lambda$.
\label{fig:tau1234quadrant}}
\end{figure}

The equations $\alpha_i=\pm{}1$ can, although the characteristic
polynomial (\ref{eq:charpoly}) itself is of degree 5, 
be solved for one variable (giving large expressions).
The complete set of lines is shown in Figure~\ref{fig:jury_tau_1}
for
\typeout{CHANGED} %%%  $\tau=1$ to
$\tau=4$ and illustrates the redundancy of the inequalities
generated by the Jury table.
Only four (three for $\tau=1$) of the $2n$ inequalities constitute the
border of stability, and unfortunately it seems one has to select 
them by hand.
\typeout{CHANGED} %%%  , giving the diagrams in Figure~\ref{fig:tau1234cut}.
Control is only necessary for $|\lambda|>1$, and folding the relevant
stability area into the same quadrant gives Fig.~\ref{fig:tau1234quadrant}
showing how $\lambda_{\rm max}$ decreases for increasing~$\tau$.

\begin{figure}[htbp]
\noindent
\begin{center}
\epsfig{file=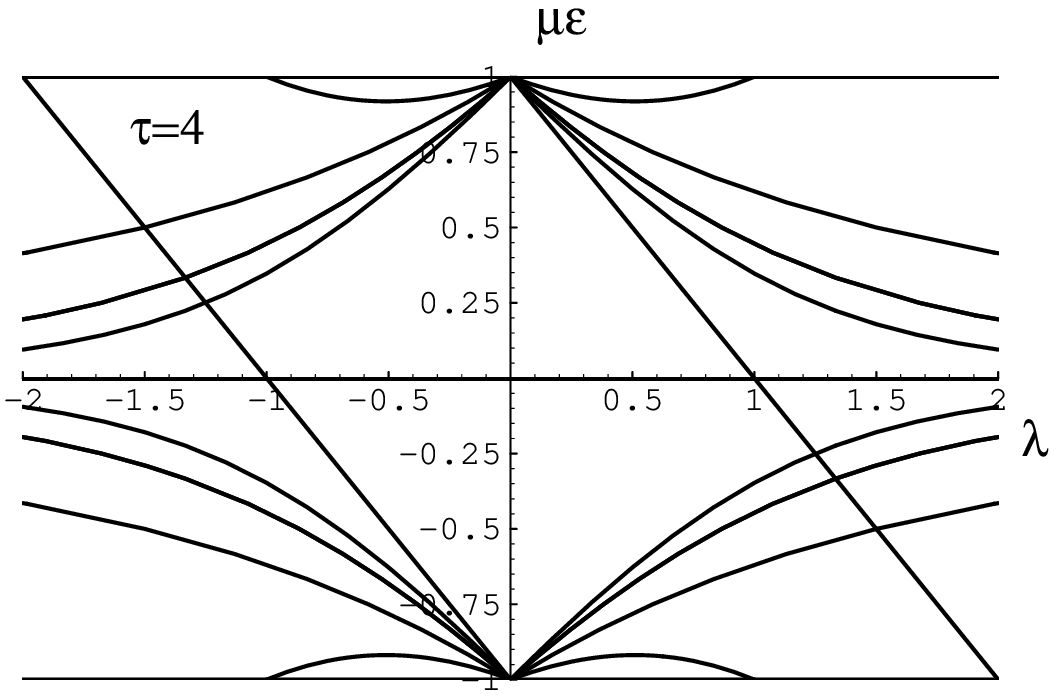, angle=270,width=0.8\columnwidth}
\end{center}
\caption[alpha1]{
Complete Jury diagram
% (complete set of lines)
 for $\tau=4$ 
% and $-2\leq\lambda\leq{}+2$ $-1\leq\mu\varepsilon\leq +1$
(see Appendix~\ref{appj1}).
\label{fig:jury_tau_1}}
\end{figure}

%\clearpage

%\setcounter{figure}{3}

\IGN{
\begin{figure}[htbp]
\noindent
\begin{center}
\epsfig{file=jury_tau_1.eps,  angle=270,width=2.15in}
\\
\epsfig{file=jury_tau_2.eps,  angle=270,width=2.15in}
\\
\epsfig{file=jury_tau_3.eps,  angle=270,width=2.15in}
\\\epsfig{file=tauwahnsinn.eps, angle=270,width=2.15in}
\end{center}
\caption[alpha1]{
 Jury diagram (complete set of lines)
 for $\tau=1, 2, 3, 4$ and
$-2\leq\lambda\leq{}+2$
$-1\leq\mu\varepsilon\leq +1$.
\label{fig:jury_tau_1}}
\end{figure}
% \clearpage
\begin{figure}[htbp]
\noindent
\begin{center}
\epsfig{file=tau1cut.eps, angle=0,width=2.15in}
\\
\epsfig{file=tau2cut.eps, angle=0,width=2.15in}     
\\
\epsfig{file=tau3cut.eps, angle=270,width=2.15in}     
\\
\epsfig{file=tau4cut.eps, angle=270,width=2.15in}     
\end{center}
\caption{
 Jury diagram
 for $\tau=1, 2, 3, 4$ and
$-2\leq\lambda\leq{}+2$
$-1\leq\mu\varepsilon\leq +1$. 
\label{fig:tau1234cut}}
\end{figure}

}%IGN

%\clearpage

%\clearpage

\section{Stabilizing unknown fixed points}
%\section{Stabilization of unknown fixed points}
\label{sec:diffkontr}
As the OGY approach discussed above requires the knowledge of the 
position of the fixed point, one may wish to stabilize 
purely by feeding back differences of the system variable 
at different times.
This becomes relevant in the case of parameter drifts
\cite{claussen98a} which often can occur in experimental
situations.
A time-continous strategy has been introduced by
Pyragas \cite{pyragas92}, and the time-discrete counterpart
(i.e.\ control amplitudes are calculated
every Poincar\'e section)
is defined by the difference control scheme \cite{bielawski93a}.
For control without delay, a simple difference control 
strategy \cite{bielawski93a,claussenthesis,schusterstemmler}
is possible for 
$\varepsilon\mu=-\lambda/3$, and eigenvalues of modulus smaller 
than unity of the matrix
 {\small $ \left( \begin{array}{cc}
 \lambda +\varepsilon\mu & - \varepsilon\mu  \\ 1 & 0
 \end{array} \right) $ \normalsize}
are obtained only for $-3 < \lambda < +1 $, so this method stabilizes 
only for oscillatory repulsive fixed points with $-3 < \lambda < -1$
\cite{bielawski93a,claussenthesis,schusterstemmler}.

%\clearpage

We can proceed in a similar fashion as for OGY control.
In the presence of $\tau$ steps delay 
the linearized dynamics of
a simple difference feedback
$r_t  = \varepsilon (x_{t-\tau}-x_{t-\tau-1})$
is given by
\begin{eqnarray}
%%%%\vec{y}_{t+1} 
\left(
\begin{array}{c}
x_{t+1}\\ \vdots\\\\\\\\\\\vdots\\x_{t-\tau}
\end{array}\right)
= 
\left(
\begin{array}{ccccccc}
\lambda&0&\cdots&&0&\varepsilon\mu&-\varepsilon\mu\\
1&0&&&&&0\\
0&1&\ddots&&&&\vdots\\
\vdots&&\ddots&&&&\\
&&&&\ddots&&\\
\vdots&&&&\ddots&0&\vdots\\
0&\cdots&&\cdots&0&1&0
\end{array}
\right)
%\vec{y}_{t}
\left(
\begin{array}{c}
x_{t}\\ \vdots\\\\\\\\\\\vdots\\x_{t-\tau-1}
\end{array}\right)
\nonumber
\end{eqnarray}
in delayed coordinates
$(x_t,x_{t-1},\ldots x_{t-\tau-1})$,
and the characteristic polynomial is given by
\begin{eqnarray}
0=(\alpha-\lambda)\alpha^{\tau + 1}+ (1-\alpha) \varepsilon\mu.
\end{eqnarray}

\noindent
As we have to use $x_{t-\tau-1}$  in addition to 
$x_{t-\tau}$, the system is of dimension $\tau+2$,
and the lower bound of Lyapunov numbers that can be controlled
are found to be 
\begin{eqnarray}
\lambda_{\rm{}inf} = - \frac{3+2\tau}{1+2\tau} 
=- \left(1+\frac{1}{\tau+ {1}/{2} } \right)
\label{eq:lambdamaxdiff}
\end{eqnarray}
and the asymptotic control amplitude at this point is 
% given by
\begin{eqnarray}
\varepsilon\mu = \frac{ (-1)^{\tau} }{1+2\tau}.
\end{eqnarray}

%\setcounter{figure}{5}

% omit in new version...
\IGN{
\begin{figure}[htbp]
\noindent
\begin{center}
\epsfig{file=jurydiff_tau_0_cut.eps, angle=0, width=2.1in}
\\
\epsfig{file=jurydiff_tau_1_cut.eps, angle=0, width=2.1in}
%\end{center}
%\begin{center}
%%\epsfig{file=jurydiff_tau_1_cut.eps, angle=0, width=2.5in}
%
\\
\epsfig{file=jurydiff_tau_2.eps, angle=0, width=2.1in}
\\
\epsfig{file=jurydiff_tau_right3cut.eps, angle=0, width=2.1in}
%alle Linien tau=3:
%\epsfig{file=jurydiff_tau3_alle_richtig.eps, angle=0, width=2.5in}
\end{center}
\caption[Jury difference tau 0]{ 
Difference feedback for $\tau=0$, $\tau=1$, $\tau=2$, and $\tau=3$:
The lines derived by the Jury criterion define the borders of
the stability area.
%%%%%%%%%%(ordinata: $\lambda$, abscissa: $\mu\varepsilon$).
The case $\tau=0$ has already been given by 
\cite{bielawski93a}.
\label{fig:jurydiff0neben1}}
\end{figure}
}%

%very old:
\IGN{
\begin{figure}[htbp]
\begin{center}
\epsfig{file=jurydiff_tau_0123.eps, angle=0, width=5in}
\end{center}
\caption[Jury difference tau 01 combined]{
Difference feedback for $\tau=0,1$:
Lines derived by the Jury criterion, they give the border of
the stability area.
(ordinata: $\lambda$, abscissa: $\mu\varepsilon\cdot(-1)^{\tau}$)
\label{fig:jurydiff01}}
\end{figure}
}

%\clearpage

% Combined:::::

\begin{figure}[htbp]
\noindent
\epsfig{file=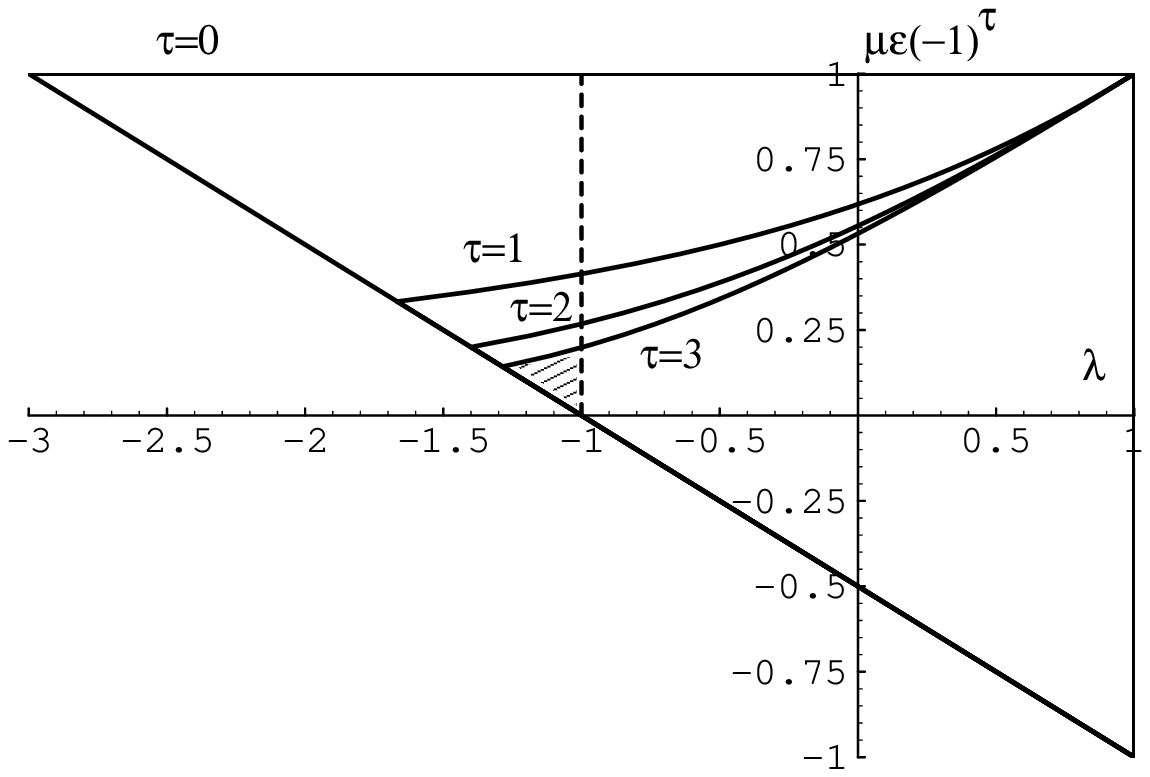, angle=0 ,width=0.9\columnwidth}
\caption[Jury difference tau 01 combined]{
Difference feedback for $\tau=0,1,2,3$:
Stability borders derived by the Jury criterion
(see Appendix~\ref{appj2}).
The stability diagram of the non-delayed case $\tau=0$ has already been given 
in \cite{bielawski93a}.
From $\lambda=-1$ (dashed line) to $\lambda=+1$ the
system ist stable without control.
For each $\tau$, 
control is effective only within the respective 
area (shaded for $\tau=3$). 
%%% (ordinata: $\lambda$, abscissa: $\mu\varepsilon\cdot(-1)^{\tau}$)
\label{fig:jurydiff01}}
\end{figure}

%\clearpage

The stability area in the $(\mu\varepsilon,\lambda)$ plane 
is bounded by the lines $\alpha_i=\pm{}1$ 
where $\alpha_i$ are the coefficients given by the 
Jury criterion \cite{jury} (see Appendix~\ref{appj0}).
For $\tau=0$, the Jury coefficients are
$
\alpha_1 =- \frac{\lambda+\varepsilon\mu}{1+\varepsilon\mu}
$
and
$
\alpha_2 = \varepsilon\mu.
$
\IGN{
\begin{eqnarray}
\alpha_2 = \varepsilon\mu 
~~~~~~~~
%\mbox{and}
~~~~~~~~
\alpha_1 = 
- \frac{\lambda+\varepsilon\mu}{1+\varepsilon\mu}
\nonumber
\end{eqnarray}
}%IGN
For $\tau=1$ to $\tau=3$, the Jury coefficients are
shown in  Appendix~\ref{appj2}.

The controllable range is smaller than for
the unmodified OGY method, and is restricted
to oscillatory repulsive fixed points with 
$\lambda_{\rm{}inf} < \lambda \leq -1$.
A striking observation is that inserting $\tau+\frac{1}{2}$ for $\tau$
in eq.~(\ref{eq:lambdamaxogy}) exactly leads to the expression in
eq.~(\ref{eq:lambdamaxdiff}) which reflects the fact that the difference
feedback control can be interpreted as a discrete first derivative, taken at
time $t-(\tau+\frac{1}{2})$.
Thus the controllability relation~(\ref{eq:controllability})
holds again.

As $\lambda^{-1}$ is implying a natural time scale 
(that of exponential separation) of an orbit, it is
quite natural that control becomes delimited
by a border proportional to a product of $\lambda$
and a feedback delay time. 
Already  without the additional
difficulty of a measurement delay
this is expected to appear 
for any control scheme
that itself 
is using time-delayed 
feedback:
E.g.\ the extensions 
of time-discrete control schemes discussed in \cite{socolar98}
with an inherent Lyapunov number limitation
due to memory terms,
and the experimentally widely applied 
time-continuous schemes 
Pyragas and ETDAS
\cite{just97,just99pla,franc99}.
Here Pyragas control has 
the Lyapunov exponent limitation $\Lambda\tau_{\sf p}\leq{}2$ 
together with the requirement of the Floquet multiplier 
of the uncontrolled orbit having an imaginary part of
$\pi$, meaning that deviations from the orbit after one
period experience to be flipped around the orbit by that angle,
which is quite the generic case \cite{just99pre}.
This nicely corresponds with the requirement
of a negative Lyapunov number that appears in difference control.
A positive Lyapunov number in the time-discrete picture
corresponds to a zero flip of the time-continuous orbit,
and is consistently uncontrollable in both schemes.

Recently, the influence of a control loop latency 
has also been studied for continuous time-delayed feedback
\cite{just99pre} by Floquet analysis,
obtaining a critical value
%  of $\tau(1-\lambda \tau_{\sf p}/2)/(\lambda \tau_{\sf p})$ 
for the measurement delay $\tau$, 
that corresponds to a maximal Lyapunov exponent
$\exp(\lambda_{\sf inf})= \Lambda\tau_{\sf p} = \frac{1}{1/2 + \tau/\tau_{\sf p}}$,
where $\tau_{\sf p}$ is the orbit length and matched feedback delay.
By the log inequality that again translates (for small Ljapunov exponents)
to our result for the time-discrete
difference control.
An exact coincidence could not be expected, as 
in Pyragas control the feedback difference is computed
continuously sliding with the motion along the orbit,
where in difference control it is evaluated within each
Poincar\'e section.
Although the time-continuous case (as an a priori infinite-dimensional
delay-differential system) could exhibit much more
complex behaviour, it however astonishing that for all three methods,
OGY, difference, and Pyragas control,
the influence of measurement delay mainly results in the same
limitation of the controllable Lyapunov number.

%\clearpage

%\vspace{4in}

\section{Conclusions}
Delayed measurement is a generic problem that can appear in
controlling chaos experiments.
In some situations it may be technically impossible to extend the
control method, then one wants to know the stability borders
with minimal knowledge of the system.

We have shown that both OGY control and difference control cannot
control orbits with an arbitrary Lyapunov number if there is
only delayed knowledge of the system.
The maximal Lyapunov number 
up to which an instable orbit can be controlled
is given by $1+\frac{1}{\tau}$ for 
OGY control and $1+{1}/{(\tau+{1}/{2})}$ for difference control.
For small $\tau$ the stability borders can be derived by the
Jury criterion, so that the range of values for
the control gain $\varepsilon$ can be determined from the
knowledge of the Taylor coefficients $\lambda$ and $\mu$.
If one wants to overcome these limitations, 
one has to modify the control strategy.

This can be done either by applying control rhythmically
\cite{bielawski93a,claussen98c,claussenthesis,schusterstemmler} 
being equivalent 
in a formal sense to controlling the $\tau+1$-fold iterate
(for OGY control) resp. $\tau+2$-fold iterate
(for difference control) 
of the original system
\cite{claussen98c,claussenthesis}.
However, for larger values of $\tau$, 
the required values for the control gain
grow exponentially with $\tau$, because the possibility
of applying control in the intermediate time steps 
is not used.

The other possibility to improve control are memory methods
\cite{claussen98c,claussenthesis,claussen98a,socolar98}
For negative $\lambda$, in the non-delayed case $\tau=0$ the
stability area can be extended
by two methods including an averaged resp.\ decaying
memory and requiring only one extra parameter \cite{socolar98}.

For arbitrary
% but known 
$\lambda$ and delayed measurement, 
an improved control 
even ensuring all eigenvalues to become zero
can be acheived by memory methods
that take into account control amplitudes applied in previous time
steps \cite{claussen98c,claussenthesis}.

\typeout{CHANGED}
\IGN{
However, in some situation it may be not feasible to apply
those memory methods, e.g.
if it is not possible to change an experimental setup.
There may be also the situation where the exact delay length
is not known, and only certain orbits are accessible
for control. In these cases our
control borders 
\ref{eq:lambdamaxogy}
for OGY control and
\ref{eq:lambdamaxdiff}
for difference control 
delimit the maximal control area as a function 
of the delay $\tau$.
}%

%\clearpage

\appendix

\section{The Jury criterion}
\label{appj0}
\noindent
The Jury criterion \cite{jury} gives a sufficient and necessary condition
that all roots of a given polynomial are of modulus smaller than 
unity.
Given a polonomial 
%\begin{eqnarray}
$
P(x)=
a_n x^n + a_{n-1} x^{n-1} + \cdots + a_1 x + a_0, 
$
%\nonumber\end{eqnarray}
one applies the iterative scheme of the {\it Jury table:}
\begin{eqnarray}
\forall_{ 0 \leq i \leq n}
\;\;\;\;
b_i &:=& a_{n-i} 
\nonumber
\\
\alpha_n &:= & b_n/a_n 
\nonumber
\\
\forall_{ 1 \leq i \leq n}
\;\;\;\;
a_{i-1}^{\rm new} 
&:=& a_i - \alpha_n b_i
\nonumber
\end{eqnarray}
giving $\alpha_n$ and
coefficients $a_{n-1}{\ldots}a_0$ for the next iteration.

The Jury criterion states that the eigenvalues are of modulus smaller 
than unity if and only if 
%\begin{eqnarray}
$
\forall_{ 1 \leq i \leq n}
%% \;\;
|\alpha_i| < 1.
$
%\nonumber \end{eqnarray}  

\noindent
The criterion gives $2n$ (usually partly redundant)
inequalities that define hypersurfaces 
in coefficient space. 
These hypersurfaces are given by algebraic equations;
%% , and 
it is not necessary 
to compute the roots of the polonomial.

Whereas the Jury criterion is extremely helpful for small $n$ and for
numeric purposes, the hypersurface equations become very complex for 
large~$n$, and one has to select the relevant hypersurface equations.
~~
---
~~
{\sl 
Two additional necessary conditions 
(`check-first conditions')
for stability are
%\begin{eqnarray}
$(-1)^n a_n P(-1) >0
$ and
$       a_n P(+1) >0$.
}
%\nonumber\end{eqnarray}
%
\normalsize
\clearpage
%\onecolumn
%\begin{widetext}
\onecolumngrid
\section{Jury coefficients for unmodified OGY control}
\label{appj1}
\vspace*{-0.3cm}
\noindent
For $\tau=2$, the Jury coefficients take the values
 $\displaystyle
\alpha_3 =  -\mu\varepsilon,
~~
\alpha_2 =  -\frac{\lambda\mu\varepsilon}{1-(\mu\varepsilon)^2},
~~
\alpha_1 = -\frac{\lambda}{
1-(\mu\varepsilon)^2 + \lambda \mu\varepsilon}
$\\
and among the crossing points of the six lines given by $\alpha_i=\pm{}1$
one finds the maximal Lyapunov number $\lambda=\pm{}3/2$.
\\
For $\tau=3$ the Jury coefficients are:
\begin{eqnarray}\nonumber
\alpha_4 = -\mu\varepsilon,
~~~~
\alpha_3 = -{{\lambda \mu\varepsilon}\over {1 - {(\mu\varepsilon)^2}}},
~~~~
\alpha_2 = {{{\lambda^2} \mu\varepsilon}\over {-1 + 2\,{(\mu\varepsilon)^2} 
+ {\lambda^2}\,{(\mu\varepsilon)^2} - {(\mu\varepsilon)^4}}},
~~~~
\alpha_1 = {{\lambda - \lambda
(\mu\varepsilon)}\over {-1 + \mu\varepsilon + {\lambda^2} \mu\varepsilon 
+ {(\mu\varepsilon)^2} - {(\mu\varepsilon)^3}}}
\end{eqnarray}
For $\tau=4$, the borders given by the Jury coefficients
are already described by algebraic equations of higher order:
%\small
\begin{eqnarray}
\alpha_5 &=& -\mu\varepsilon,
~~~~~~~~
\alpha_4 ={{\lambda\,\mu\varepsilon}\over {-1 + {(\mu\varepsilon)^2}}},
~~~~~~~~
\alpha_1 = 
 {{\lambda\,\left( 1 - \lambda\,\mu\varepsilon - {(\mu\varepsilon)^2} \right) }\over 
    {-1 + \lambda\,\mu\varepsilon + {\lambda^3}\,\mu\varepsilon + 2\,{(\mu\varepsilon)^2} 
+ {\lambda^2}\,{(\mu\varepsilon)^2} - \lambda\,{(\mu\varepsilon)^3} - {(\mu\varepsilon)^4}}}
\nonumber
\\
\alpha_3 &=&
{{{\lambda^2}\,\mu\varepsilon}\over {-1 + 2\,{\mu\varepsilon^2} 
+ {\lambda^2}\,{\mu\varepsilon^2} - {(\mu\varepsilon)^4}}},
~~~~~~~~
\alpha_2 =
{{{\lambda^3}\,\mu\varepsilon}\over 
    {-1 + 3\,{(\mu\varepsilon)^2} + 2\,{\lambda^2}\,{(\mu\varepsilon)^2} 
+ {\lambda^4}\,{(\mu\varepsilon)^2} - 3\,{(\mu\varepsilon)^4} - 
      2\,{\lambda^2}\,{(\mu\varepsilon)^4} + {(\mu\varepsilon)^6}}}
\nonumber
\end{eqnarray}
\normalsize
The equations $\alpha_i=\pm{}1$ can, although the polonomial is of degree 5, 
be solved for one variable,
see Fig.~\ref{fig:jury_tau_1}.
% (giving large expressions).
%The complete set of lines are shown in Figure~\ref{fig:jury_tau_1}.
%
%
\vspace*{-0.3cm}
\section{Jury coefficients for unmodified difference control}
\label{appj2}
\vspace*{-0.3cm}
\noindent
For $\tau=1$, the Jury coefficients are
 $\displaystyle  
\alpha_3 = \varepsilon\mu,
~~
\alpha_2 = \varepsilon\mu (\lambda-1) (1-(\varepsilon\mu)^2)^{-1},
~~
\alpha_1 =
\frac{
((\varepsilon\mu)^2 - \lambda)(1-(\varepsilon\mu)^2 -
(\lambda-1)\varepsilon\mu
}{
(1-(\varepsilon\mu)^2)^2 - (\varepsilon\mu)^2 (\lambda-1)^2 }
$\\
For $\tau=2$, the Jury coefficients are
%sheet_abcde und a=1 b=-l c=0 d=-m e=m
%alpha1  ((1 + m)*(-l + m^2))/(1 + m - l*m + l^2*m - 2*m^2 + l*m^2 - m^3)
%alpha2  ((1 - l)*m*(l - m^2))/(-1 + 3*m^2 - 2*l*m^2 + l^2*m^2 - m^4)
%alpha3  ((-1 + l)*m)/(1 - m^2)
%alpha4  m
%\small
\begin{eqnarray}
\alpha_4 &=& \varepsilon\mu,
\nonumber
~~~~
\alpha_3 =  {{\left( -1 + \lambda \right) \varepsilon\mu}\over 
{1 - {(\varepsilon\mu)^2}}}
~~~~
\alpha_2 = {{\left( 1 - \lambda \right) \varepsilon\mu
 \left( \lambda - {(\varepsilon\mu)^2} \right) }\over 
    {-1 + 3 {(\varepsilon\mu)^2} - 2 \lambda {(\varepsilon\mu)^2} 
+ {\lambda^2} {(\varepsilon\mu)^2} - {(\varepsilon\mu)^4}}}
\\ \nonumber
\alpha_1 &=&  {{\left( 1 + \varepsilon\mu \right) 
\left( -\lambda + {(\varepsilon\mu)^2} \right) }\over 
    {1 + \varepsilon\mu - \lambda \varepsilon\mu + 
{\lambda^2} \varepsilon\mu - 2 {(\varepsilon\mu)^2} 
+ \lambda {(\varepsilon\mu)^2} - {(\varepsilon\mu)^3}}}
\nonumber     
\end{eqnarray}
\normalsize
%\tiny
For $\tau=3$, the Jury coefficients are
\begin{eqnarray}
\alpha_5 &=& \varepsilon\mu,
~~~~~~~~
\alpha_4 =  {{\varepsilon\mu - \lambda \varepsilon\mu}\over 
{-1 + {(\varepsilon\mu)^2}}},
\nonumber
~~~~~~~~
\alpha_3 = {{\left( 1 - \lambda \right) \varepsilon\mu
\left( \lambda - {(\varepsilon\mu)^2} \right) }\over
    {-1 + 3 {(\varepsilon\mu)^2} - 2 \lambda {(\varepsilon\mu)^2} 
+ {\lambda^2} {(\varepsilon\mu)^2} - {(\varepsilon\mu)^4}}}
\nonumber
\\
\alpha_2 &=& {{\left( 1 - \lambda \right) \varepsilon\mu
{{\left( \lambda - {(\varepsilon\mu)^2} \right)
}^2}}\over
    {-1 + 5 {(\varepsilon\mu)^2} - 4  \lambda {(\varepsilon\mu)^2} 
+ 3{\lambda^2} {(\varepsilon\mu)^2} 
- 2{\lambda^3} {(\varepsilon\mu)^2} +   
      {\lambda^4} {(\varepsilon\mu)^2} 
- 3 {\lambda^2} {(\varepsilon\mu)^4} + {(\varepsilon\mu)^6} 
- 36 \lambda {(\varepsilon\mu)^8}}}
\nonumber
\\
\alpha_1 &=& {{\lambda - \lambda \varepsilon\mu
+ {\lambda^2} \varepsilon\mu - {(\varepsilon\mu)^2} - \lambda {(\varepsilon\mu)^2} 
+ {(\varepsilon\mu)^3} - \lambda {(\varepsilon\mu)^3} +
{(\varepsilon\mu)^4}}\over
    {-1 + \varepsilon\mu - \lambda \varepsilon\mu + {\lambda^2}
\varepsilon\mu 
- {\lambda^3} \varepsilon\mu 
+ 3 {(\varepsilon\mu)^2} - 2  \lambda {(\varepsilon\mu)^2} +
      {\lambda^2} {(\varepsilon\mu)^2} - 2 \lambda {(\varepsilon\mu)^3} 
- {(\varepsilon\mu)^4}}}
\nonumber     
\end{eqnarray}
\normalsize
\vspace*{-1.0cm}
%
%\end{widetext}
\twocolumngrid
\vspace*{-1.6cm}
%\begin{references}

\end{document}